\documentstyle[aps,twocolumn]{revtex}

\input epsf
\epsfverbosetrue
\begin{document}

\draft
\preprint{UBCTP-95-010}
\title{Singular Behaviour of Electrons and of Composite Fermions in a 
Finite Effective Field} 

\vskip .5in

\author{S. Curnoe and P. C. E. Stamp}

\vskip .5in

\address{Department of Physics, University of British Columbia,\\
Vancouver, B.C. V6T 1Z1, Canada}
\maketitle
\vskip .5in
\begin{abstract}
We calculate the self-energy $\Sigma_n(\epsilon)$ of fermions in 
Landau level $n$, in a finite field.  Two cases are considered, in 
which fermions couple either to gauge fluctuations (as in the 
composite fermion gauge theory) or to phonons, as an example of a Fermi 
liquid. 
Perturbative calculations of the composite fermion spectrum show an
unphysical suppression of the quasiparticle spectral weight at the 
composite fermion levels.  We argue that this problem might be 
resolved by a non-perturbative calculation; alternatively, the 
system might be unstable.
\end{abstract}  
\pacs{PACS Nos:}
\vskip .2in

\narrowtext

The original Laughlin Theory \cite{laugh} of the fractional quantum Hall 
effect explains the ground state and low energy excitations for filling
fractions $\nu = 1/(2k+1)$, with $k$ a positive integer.  
Recent experiments \cite{expt} indicate a more general theory is needed
to explain both the gapless state at $\nu=1/2$, and the hierarchy of states for
a general $\nu$. 
A promising candidate involves new quasiparticles  \cite{jain} called
``composite fermions" (CF's). CF's can be viewed as ordinary electrons, to 
each of which is attached an artificial flux tube, containing two flux
quanta, oriented oppositely to the applied field {\bf B}. 
The net mean field {\bf b} acting on the CF's becomes 
$b=B-\bar{B}_{1/2r}$, where 
$\bar{B}_{1/2r}=4\pi r c\rho_e/e$ is the mean field from the artificial
flux, and $\rho_e$ is the mean electron density; we see that $b=0$
when $\nu = 1/2r$.
The set of FQHE states at $\nu = p/(1+2r p)$, with $p=0,\pm 1,\pm 2$, etc.,
then arises because $b$ is such that an integer number $|p|$ of CF Landau
levels is filled; the ``principal hierarchy" has $r=1$, and, as 
$|p|\rightarrow\infty$, so $\nu\rightarrow 1/2$, the ``bare" CF gap
$\tilde{\omega_c}=eb/\tilde{m}$ goes to zero (here $\tilde{m}$ is the ``bare"
CF mass, ignoring CF interactions).

A field theory of CF's \cite{lopez} demonstrates that CF
interactions resemble the gauge interactions in the gauge theory of
high-$T_{c}$ superconductors \cite{baska},\cite{lee}, and this has led 
to several calculations of the renormalised properties of the $\nu=1/2$
state \cite{lopez} and the FQHE state with fractions near $1/2$;
the FQHE calculations have looked at the renormalised gap
\cite{lopez},\cite{kim1}, the current response \cite{simon}, and
thermodynamic properties
like the compressibility and density \cite{kim1}, in perturbative
studies of the gauge interaction.    However these gauge interactions 
have severe infra-red divergences; they contain a term \cite{lopez}
\begin{equation}
        D_{11}(q,\omega) = \frac{q}{\chi q^s - i \gamma\omega}
\end{equation}
where $3\geq s\geq 2$; if the Coulomb interactions between the CF's
are screened by bringing a conducting plate up to the 2-d semiconductor,
then $s=2$, whereas $s=3$ for the completely unscreened case.
If the CF self-energy is calculated perturbatively, to first order in 
$D_{11}(q,\omega)$, then one finds, at $\nu=1/2$, that at $T=0$
\begin{equation}
    \Sigma(p,\epsilon) \sim \left\{ \begin{array}{ll}
                             (i\Omega_0/\epsilon)^{1/3}\epsilon   &   s=3    \\
                             \epsilon\log(\epsilon)  &  s=2
                             \end{array} 
                               \right.
      \label{eq:hlr}
\end{equation}
for $p$ near the $\nu =1/2$ Fermi surface.
Non-perturbative calculations have also been done of $\Sigma(\epsilon)$
and of the response functions using eikonal expansions \cite{stamp1},
$1/N$ expansions \cite{kim2}, and renormalisation group analysis 
\cite{chakr}. A crucial feature of these, heavily emphasized in \cite{stamp1},
\cite{kim2}
and also in a recent paper of Stern and Halperin \cite{stern}, is that
Ward identities force the correlation functions to be much less
singular than the self-energy.  The self-energy is often considered to 
be unphysical since it is not gauge invariant (which causes the $\nu=1/2$
self-energy to be {\em infinite} \cite{lee} at any finite $T$).    
However, one can also argue that its {\em pole structure} is gauge invariant
and therefore physically meaningful.  This is assumed by Stern and
Halperin \cite{stern}, in their analysis of the $s=2$ case, and 
certainly one naively expects $\Sigma(\epsilon)$ to be well behaved 
away from $\nu = 1/2$,  at least if 
$\bar{\epsilon} \ll \tilde{\omega}_c$, when the unrenormalised 
gap $\tilde{\omega}_c$ cuts off the IR
divergences.

In this paper we take a closer look at this question, by calculating the
self-energy $\Sigma_n(\epsilon)$ perturbatively, for a fermion in Landau
level $n$.
This calculation is applied to both CF'S and also to an ordinary Fermi liquid
in an applied field. 
We find that $\Sigma_n(\epsilon)$ has a rather peculiar singular structure. 
A less singular structure exists even for Fermi liquids.

We start from the usual lowest order
perturbative expression for a 2-d electronic 
self-energy:
\begin{eqnarray} 
     \Sigma_{n}(\epsilon)  = & &
                   \int \frac{d^2q}{(2\pi)^2} \int_{0}^{\infty} 
                    \frac{d\omega}{\pi}
                   {\rm Im}U(q,\omega) 
                    \sum_{m=-n}^{\infty}
                     |\Lambda^{m}_{n}(q)|^{2} \nonumber \\ 
& &\times            \left(\frac{1+n_{B}(\omega)-n_{f}(n+m)}
           {\epsilon-(n+m)\tilde{\omega}_c-\omega+i\delta} \right. \nonumber\\
& & \left. + \frac{n_{B}(\omega)+n_{f}(n+m)}
                    {\epsilon -(n+m)\tilde{\omega}_c+\omega+
                       i\delta} \right)
\end{eqnarray} 
where $n,m$ are Landau level indices , $n_f(r)=n_f((r+1/2)\tilde{\omega}_c)$
is the Fermi distribution for the $r$th Landau level, $n_B$ is the Bose
distribution and $U(\omega,q)$ describes the relevant interaction 
fluctuation. For
gauge fluctuations
\begin{equation}
     U(q,\omega) \sim -4 \tilde{v}_f^2 \left( \frac{eb}{2\pi}\right)
                       \left| \frac{\vec{k}_f\times\hat{q}}{m}\right|^2
                        D_{11}(q,\omega)
\end{equation}
whereas for Fermi liquids $U(q,\omega)\sim f^2\chi(q,\omega)$, where 
$\chi(q,\omega)$ is the relevant fluctuation propagator and $f$ is the
relevant Landau parameter \cite{stamp2}.
As a concrete example we shall choose the coupled electron-phonon Fermi liquid,
for which
\begin{equation}
    U(q,\omega) \sim 2 \bar{g}^2(q v_f)^2 \frac{q c_s}{\omega^2-q^2c_s^2}
\end{equation}
where $c_s$ is the sound velocity and $\bar{g}$ a dimensionless coupling.
Finally, $\Lambda^m_n(q)$ is the overlap matrix element between plane
wave and Landau level states, given by \cite{brail}
\begin{eqnarray}
& &        |\Lambda^m_n(q)|^2 =  \nonumber \\
& & \left(\frac{q^2\mit{l}^2_0}{2}\right)^m 
                       e^{-q^2\mit{l}_0^2/2} \frac{(n+m)!}{n!}
                     |L^{n+m}_n\left(\frac{q^2\mit{l}^2_0}{2}\right)|^2
\end{eqnarray}
where $L^{n+m}_n$ is a Laguerre polynomial and 
$\mit{l}_0= (hc/eb)^{1/2}$ is the Landau length.

     We calculate $\Sigma_n(\epsilon)$ in a quasiclassical approximation
\cite{kim1},
\cite{simon}, assuming $N \gg 1$ filled Landau levels (for CF's, $N =|p|$).
The $T=0$ calculations can be done analytically, and for
 $kT \ll \tilde{\omega}_c$, one can expand about the $T=0$ answers
\cite{kim1}; notice the CF theory is only meaningful if 
$kT \ll  \tilde{\omega}_c$.

\begin{figure}
\epsfysize=4.0in
\epsfbox[60 132 585 750]{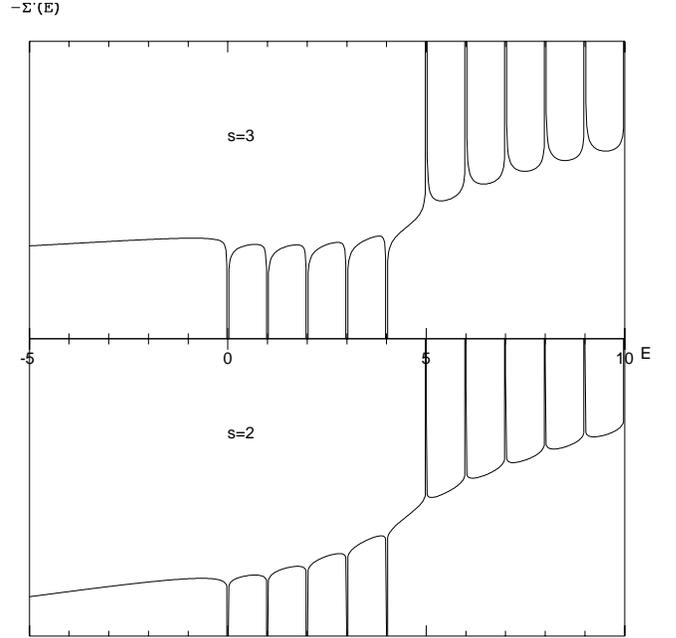}
\caption{The real part $\Sigma_n(\epsilon)$ of the self energy for
composite fermions shown for  $s=3$
and $s=2$.
We assume $p=5$ (so that $\nu =5/11$). The temperature is
$kT=.03\tilde{\omega}_c$ and the chemical potential is pinned halfway between
the fifth and sixth Landau levels.
\label{fi:rselfe}}
\end{figure}

Consider first the CF gauge theory results; writing 
$\Sigma =  \Sigma^{'}-i\Sigma^{''}$, we find
\begin{eqnarray}
 \Sigma_{n}^{'}(\bar{\epsilon})&  =&  \frac{-{\rm sgn}(\bar{\epsilon}-N)
                    \Sigma^{''}_{n}(\bar{\epsilon})}{\sqrt{3}} \nonumber \\
    &  +&\frac{2K_s}{\sqrt{3}}
 \left(\sum_{m=0}^{\min(\lfloor\bar{\epsilon},
                       \lfloor N)}(\bar{\epsilon}-m)^{-\alpha}
    \right. \nonumber \\
& &\left.   -\sum_{m=\max(\lfloor\bar{\epsilon}+1,\lfloor N+1)}^{\infty}
                         (m-\bar{\epsilon})^{-\alpha}\right) 
\end{eqnarray}
\begin{eqnarray}
& &   \Sigma_{n}^{''}(\bar{\epsilon})  =  \nonumber \\
& &           K_s  \left(\sum_{m=0}^{\lfloor\bar{\epsilon}}
              (n_{B}(\bar{\epsilon}-m)+n_{f}(N-m))(\bar{\epsilon}-m)^{-\alpha}
                           \right.   \nonumber \\
& & \left.   + \sum_{\max(\lfloor\bar{\epsilon}+1,0)}^{\infty}
            (n_{B}(m-\bar{\epsilon})+n_{f}(m-N))(m-\bar{\epsilon})^{-\alpha}
                          \right)
\end{eqnarray}
for $s > 2$  and
\begin{eqnarray}
& &\Sigma_{n}^{'}(\bar{\epsilon})  = 
                           K_2 \log|m-\bar{\epsilon}| \nonumber\\
  & &       \times    \left(-\sum_{m=0}^{\lfloor N}
                              n_{f}(m-N) 
                     +\sum_{m=\lfloor N+1}^{\infty}n_{f}(N-m)\right)
           \\
& &  \Sigma_{n}^{''}(\bar{\epsilon})  = 
                     K_2 \left(\sum_{m=0}^{\lfloor\bar{\epsilon}} 
                    (n_{B}(\bar{\epsilon}-m)+n_{f}(N-m)) \right.\nonumber \\
& &        \left.   +\sum^{\infty}_{m=\max(0,\lfloor\bar{\epsilon}+1)}
                             (n_{B}(m-\bar{\epsilon})+n_{f}(m-N)) \right)
\end{eqnarray}
for $s=2$. In these equations $K_s$ is a constant, 
$\bar{\epsilon} = \epsilon/\tilde{\omega}_c$, $\lfloor r$ is the greatest
integer less than $r$, and the exponent $\alpha = (s-2)/s$ is positive.
We show $\Sigma^{'}_n(\epsilon)$ in Fig. (\ref{fi:rselfe}) for $s=2,3$.
When $s=3$ there are inverse cube root divergences 
$|\bar{\epsilon} - r|^{-1/3}$ in $\Sigma^{'}(\bar{\epsilon})$ as one
approaches each Landau level, from either side. 
 When $s=2$ these become logarithmic divergences.
However $\Sigma^{''}_{n}(\epsilon)$ only shows divergences, for $s>2$,
when $\epsilon \rightarrow r\omega_c$ from {\em above}; it is finite if
$\bar{\epsilon}= r + 0^{-}$.
As $\tilde{\omega}_c \rightarrow 0$ the strength of these divergences 
vanishes and we are left with the smooth curves of Eq. (\ref{eq:hlr}) above.

An apparent pathology in these results is seen by calculating
 $z^{-1}_n(\epsilon) = 1-\partial\Sigma^{'}_n/\partial\epsilon$; one then 
sees that as $\epsilon$ approaches $r\tilde{\omega}_c$, we get the 
divergence
$z^{-1}_n(\bar{\epsilon}) \sim \pm |\bar{\epsilon}-r|^{-(1+\alpha)}$,
ie., an inverse power divergence.
That the wave-function renormalisation $z^{-1}_n$  should diverge to
$+\infty$ might have been expected;
but it also shows an unnerving divergence to $-\infty$ near each Landau
level.

\begin{figure}[h]
\epsfysize=3.3in
\epsfbox[-10 132 575 700]{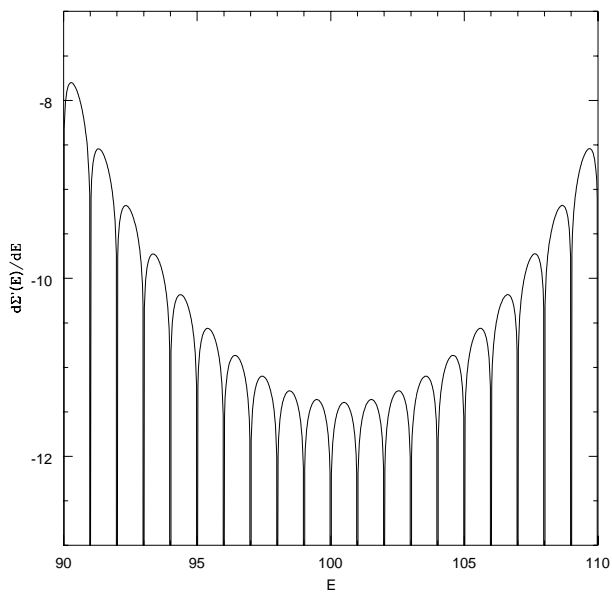}
\caption{The derivative $\partial\Sigma^{'}(\epsilon)/\partial\epsilon$ for
a Fermi liquid (here an electron phonon system) with 100 filled Landau
levels, calculated at $T=0$.
\label{fi:Drselfefl}}
\end{figure}

Notice that a  divergence in $|z^{-1}_n|$ is not unique to these singular
interactions, only the sign.
An equivalent calculation for the electron-phonon
problem gives
\pagebreak
\begin{eqnarray}
   \Sigma_{n}^{'}(\bar{\epsilon})& = & 
              \frac{K_{\phi}}{\pi}\left(
                 \sum_{m=\lfloor N+1}^{\infty}(m-\bar{\epsilon})\log\left|
                  \frac{\omega_{D}+m-\bar{\epsilon}}{m-\bar{\epsilon}}\right|
         \right. \nonumber\\
& &\left.     +\sum_{m=0}^{\lfloor N}(\bar{\epsilon}-m)\log\left|
            \frac{m-\bar{\epsilon}}{m-\bar{\epsilon}-\omega_{D}}\right|\right)
     \label{eq:rselfefl}
\end{eqnarray}
at $T=0$, with a trivial generalisation to finite $T$.   Here $K_{\phi}$ is
a constant and $\omega_D$ is a Debye cut-off (ie.,
$\omega_D = \theta_D/\tilde{\omega}_c)$. 
In Fig. \ref{fi:Drselfefl} we plot $\partial \Sigma^{'}/\partial\epsilon$,
 derived from
(\ref{eq:rselfefl}); the divergences now have the form
$z^{-1}_n(\epsilon) \sim - \log|\bar{\epsilon}-r|$,
and we still get regions with very large positive $z^{-1}_n(\epsilon)$.
This result clearly has nothing to do with any lack of gauge invariance
of $\Sigma_n(\epsilon)$.

The mathematical origin of these results
 is as follows.  First,  we assumed 
a quasi-particle form for the internal fermion line in $\Sigma_n(\epsilon)$;
in fact we assumed a form, for a system of unit area, given by 
\begin{equation}
   l_0^2 G_n(\epsilon) = \frac{1}{\epsilon-n\tilde{\omega}_c-\Sigma_n(\epsilon)}
                     \sim \frac{z}{\epsilon-\tilde{n}\omega_c}
            \label{eq:G}
\end{equation}
with $z$ a constant renormalisation factor.
(Note that $G_n(\epsilon)\sim l_0^2$, in (\ref{eq:G}), because each Landau
level has degeneracy $l_0^{-2}$).
Second, our vertex $\Lambda^m_n$ has no structure arising from the
interactions, since we are working in lowest-order perturbation theory.
This explains the divergence in $|z^{-1}|$, for
both theories- it comes from the massive degeneracy $l_0^{-2}$ in 
each Landau level.   From this point of view the positive divergence 
of $z^{-1}$ in the electron-phonon case is no different in principle from 
what occurs for an Einstein phonon spectrum 
$\omega_q = \delta(\omega-\Omega_0)$.

On the other hand the novelty of the gauge theory result is that 
$\Sigma_n^{'}(\epsilon)$ shows a positive divergence (for 
$\epsilon > n$) on both sides of the Landau level - thereby causing
both positive and negative divergences in $z^{-1}$.
Mathematically this arises because $\Sigma^{''}_n(\epsilon)$ is
essentially composed of a set of asymmetric peaks around 
$\bar{\epsilon} = r$, of the form 
$\sim (\bar{\epsilon}-r)^{-\alpha}\theta(\bar{\epsilon}-r)$,
which have long tails for 
$(\bar{\epsilon} - r) \gg 1$.
The Hilbert transform of such a function does {\em not} change sign as 
$\bar{\epsilon}$ crosses $r$, unless $\alpha > 1/2$; for the CF gauge
theory $0\leq \alpha < 1/3$.  Thus the sign change in $z_n(\epsilon)$,
each time $\epsilon$ crosses a Landau level energy $r\omega_c$, 
is caused by these long tails.

The consequences for the physical pole structure are interesting.
The fermion spectral function $A_n(\epsilon)$, defined as usual in
terms of $G_n(\epsilon)$ by 
\begin{equation}
    G_n(\epsilon) = \int _{-\infty}^{\infty} dx 
                    \frac{A_n(x)}{\epsilon - x+i\delta_x}
\end{equation}
(where $\delta_x = \delta {\rm sgn} x$ and $\delta = 0^{+}$) goes to 
zero as $|\epsilon - r \omega_c| \rightarrow 0$; eg., for $s>2$ ($\alpha >0$),
one has
\begin{equation}
   A_n(\epsilon) \rightarrow
                 \left\{ \begin{array}{ll}
                 \frac{6}{5\sqrt{3}}(\bar{\epsilon}-r)^{\alpha}
                      &(\bar{\epsilon} >r)\\
                 c_r (r - \bar{\epsilon})^{2\alpha}
                     &  (\bar{\epsilon} <r)
                 \end{array}
                 \right\}, \mbox{   as }|\bar{\epsilon}-r|\rightarrow 0 
      \label{eq:A}
\end{equation} 
where $c_r$ is a constant.  
This structure is both unphysical and in obvious contradiction with the 
form (\ref{eq:G}), assumed for the internal fermion lines - it makes 
no sense for the spectral weight to vanish on the Landau levels.
One may also define a renormalised quasiparticle spectrum $E_n$, given
by
\begin{equation}
    E_n = n\tilde{\omega}_c - \Sigma_n^{'}(E_n)
\end{equation}
This shows a rather complicated structure when one is well away from the
Fermi surface (ie., $|n-p| \ll 1$), but for 
$|E_n| <\tilde{\omega}_c$ one finds a unique pole, which can be used to 
define a quasiparticle gap $\Delta$ which agrees with that found previously
\cite{kim1},\cite{stern}.  However the unphysical nature of
$A_n(\epsilon)$ in (\ref{eq:A}), and its inconsistency with 
(\ref{eq:G}), brings these results into question.

One's first reaction to this, in the context of singular interactions,
is that the theory is intrinsically IR divergent anyway, and that the way to 
cure divergences is to do a self-consistent calculation (just as in the
zero field case).
There have been numerous investigations of vertex corrections in the case of 
zero applied field (no Landau levels), for both the gauge theories 
\cite{kim1},\cite{stamp1}-\cite{stern} and the electron-phonon
problem \cite{grimv}.
However, as far as we know, no equivalent investigations have 
been done for the finite 
field problem, and in fact it is not completely obvious  to us how the 
finite-field consistency problem can be resolved.  On physical grounds
we expect that some remnant of the sharp Landau level structure
should survive interactions, ie., that
\begin{equation}
   l_0^2 G_n(\epsilon) \sim \frac{z_n}{\epsilon- E_n} + {\rm incoherent}
\end{equation}
near the renormalised energies $E_n$  (otherwise we would not get a 
fractional quantum Hall effect).   However as we have just seen the
large degeneracy at $ \epsilon = E_n$ will tend to cause divergences in 
$z_n^{-1}$.  The way to eliminate these is clearly through divergent vertex
corrections to the 3-point vertex $\Lambda$, to cancel those in 
$z_n^{-1}$.  An attempt to devise a self-consistent scheme of this kind is
complicated by the IR divergences that already exist in the theory as
$\tilde{\omega}_c \rightarrow 0$, and it appears as though a non-perturbative
formulation of the problem is necessary.

Another possibility is that there really is some kind of instability in 
the theory.  In experiments 
one might expect this instability to be eliminated by impurity
scattering, which leads to ``Dingle broadening" of the Landau levels,
thereby destroying the sharp Landau level structure which leads to the 
divergence in $|z^{-1}(\epsilon)|$. 
In most experiments the Dingle temperature, $T_D$, which parametrizes 
this broadening, is greater than 100 mK, and the ratio 
$2 \pi k T_D/\tilde{\omega}_c$ is rarely less than 0.1. Thus
the very narrow divergent behaviour will be very difficult to see.
However we should note that we do not 
yet have a theory which combines the effects of gauge interactions and
impurity scattering on the CF's, apart from perturbative results \cite{lopez}.
Although the experiments \cite{expt} are presently being 
analysed in terms of conventional dHvA or  SdH expressions 
involving $T_D$ (or a relaxation time), it is likely that a more 
realistic theory would involve at least an energy-dependent relaxation
time, depending very rapidly on $\epsilon$ in the 
vicinity of the Landau levels.
 
This work was supported in part by the Natural Sciences and
Engineering Research Council
of Canada. We also thank Igor Herbut, and one of us (PCES) would like 
to thank Prof. B. I. Halperin, for many useful discussions.

\end{document}